\documentclass[12pt,EJP]{iopart}
\usepackage{graphicx}
\usepackage[utf8]{inputenc}
\usepackage{amsfonts}
\usepackage{amssymb}
\usepackage{amstext}
\newcommand{\eqref}[1]{(\ref{#1})}
\newcommand{\url}[1]{\texttt{#1}}

\begin{document}
\title[Physics holo.lab learning experience for heat conduction]{Physics holo.lab learning experience: \\Using Smartglasses for Augmented Reality labwork to foster the concepts of heat conduction}
\author{M.\,P.\,Strzys$^1$, S.\,Kapp$^1$, M.\,Thees$^1$, P. Klein$^1$, P.\,Lukowicz$^2$, P.\,Knierim$^3$, A.\,Schmidt$^3$ and J.\,Kuhn$^1$}
\address{$^1$ Department of Physics, Physics Education Research Group, \\University of Kaiserslautern, D--67653 Kaiserslautern, Germany}
\address{$^2$ DFKI, Embedded Intelligence Laboratory, Trippstadter Str. 122,\\ D--67663 Kaiserslautern, Germany}
\address{$^3$ LMU München, Institut für Informatik, Germany}


\begin{abstract}
Fundamental concepts of thermodynamics rely on abstract physical quantities such as energy, heat and entropy, which play an important role in the process of interpreting thermal phenomena and statistical mechanics. However, these quantities are not covered by human visual perception and since heat sensation is purely qualitative and easy to deceive, an intuitive understanding often is lacking. Today immersive technologies like head-mounted displays of the newest generation, especially HoloLens, allow for high quality augmented reality learning experiences, which can overcome this gap in human perception by presenting different representations of otherwise invisible quantities directly in the field of view of the user on the experimental apparatus, which simultaneously avoids a split attention effect. In a mixed reality (MR) scenario as presented in this paper---which we call a holo.lab---human perception can be extended to the thermal regime by presenting false-color representations of the temperature of objects as a virtual augmentation directly on the real object itself in real-time. Direct feedback to experimental actions of the users in form of different representations allows for immediate comparison to theoretical principles and predictions and therefore is supposed to intensify the theory-experiment interactions and to increase students' conceptual understanding. We tested this technology for an experiment on thermal conduction of metals in the framework of undergraduate laboratories. A pilot study with treatment and control groups ($N = 59$) showed a small positive effect of MR on students' performance measured with a standardized concept test for thermodynamics, pointing to an improvement of the understanding of the underlying physical concepts. Theses findings indicate that complex experiments could benefit even more from augmentation. This motivates us to enrich further experiments with MR. 

\end{abstract}

\submitto{\EJP}

\maketitle

\section{Introduction}

Modern digital media, such as smartphones and tablets, with their numerous sensors covering different physical quantities, have successfully been established as portable mini-labs in the last years. Today they allow to perform experiments in many fields of physics, such as mechanics, optics, acoustics and even nuclear physics \cite{Klei15,Kuhn14,Schw13,Vogt11,Klei14,Kuhn14a}. With regard to the precision of the results, these smart media have proven to be able to substitute classical measurement devices in physics teaching scenarios at school and university. However, in experiments performed with these smart media, several design principles of cognitive-affective theory of learning with media (CATLM)\cite{Maye10}, such as spatial and temporal contiguity, can only be respected with limitations: since users always have to integrate their perception of the real world experiment and the digitally processed information of their smart measurement device, this leads to a higher level of cognitive load (CL). In cognitive load theory (CLT) \cite{Swel10} and augmented cognitive load theory (aCLT)\cite{Huk09}, which similar to CATLM additionally takes into account affective variables, like motivational and emotional factors, this is also known as the split attention effect \cite{More07}. This means that the process of spatial or temporal integration of information leads to a reduction of the amount of free working memory capacity which in turn is not available for further processing of the results on a higher level as, e.g., matching of the information presented in different representations or comparison of the experimental results with theoretical implications. 

The gap in human perception as well as the lacking spatial contiguity, however, can be overcome with the help of other technologies in the field of Virtual Reality (VR) and Augmented Reality (AR) which also have quickly made progress in recent years \cite{Schm16,Sand15,Hock16} and in the meantime finally also have entered the field of education \cite{Sant14,Kuhn16}. Today, various setups based on head-mounted displays allow to embed virtual objects into the real environment, leading to a Mixed Reality (MR) experience. To classify such scenarios the so-called virtuality continuum has been introduced \cite{Milg94}, which spans form purely virtual environments on one end, to the real world without any augmentations on the other. Smartglass technology is able to completely cover this continuum leading to a true immersive virtually augmented world experience for the users. 

In such a digitally enhanced surrounding, virtual and real objects do not only co-exist, but moreover are also able to interact with each other in real-time. In a scenario like this, digital augmentations can therefore be used to enrich human perception with further senses. Today digital sensors are available for a huge number of different physical quantities---otherwise inaccessible to human perception---like temperature, electrical current, electromagnetic fields and many more. Immersive technologies with smart glasses can now be used to merge human perception of reality with digitally visualized sensor data directly in the user's field of view, obeying spatial and temporal contiguity and thereby avoiding any split attention effect. By transferring sensor data to the visual sense, e.g., by transforming it into different representations like diagrams or false-color representations, we are finally able to make the invisible visible and the not observable apparent; such a learning scenario we call a Physics \textit{holo.lab}.

The paper is structured in the following way: As a first example, we present a holo.lab version of a classical experiment on heat conduction in metals for an introductory laboratory course in thermodynamics for physics and other STEM students. In Section \ref{hololab} we first introduce the learning scenario, give a brief recap of the theory of heat conduction and finally explain the experimental setup used in the laboratory classes in more detail. 

To investigate the effects of the AR-support on the conceptual understanding of the students, we conducted a first pilot study in a quasi experimental 2$\times$2-design in the introductory physics laboratory classes at university. In Section \ref{study} we give an overview of the theoretical background and methods, present the results of this empirical study and discuss its implications. 

\section{A holo.lab for heat conduction}\label{hololab}

In this paper we present a holo.lab version of a standard experiment on heat conduction in metals for an introductory laboratory course in thermodynamics, using physical data from external sensors---in this case an infrared camera---for analyzing and displaying physical phenomena to the users. Recently, the educational potential of infrared (IR) cameras combined with the benefit of false-color representations visualizing thermal phenomena has been focused and many applications and experiments on the level of primary school up to university Physics have been developed \cite{Voll01,Moel07,Voll13,Hagl16,Hagl16a,Nord16} also including projection-based AR technologies \cite{Palm16}, where thermal images are projected directly onto the physical objects for energetic analysis. Our holo.lab setup goes one step further, using false-color representations as digital augmentation in form of HoloLens ``holograms'' placed directly at the real physical object. These augmentations are mutually 3D and registered in real space. In our setup students yield the possibility to observe the heat flux through a metallic rod, heated on one side while simultaneously cooled on the other, from all angles without the problem of occlusion. Moreover, additional representations as, e.g., graphs and numerical values, can also be included as digital augmentations to the real experiment (cf. Fig.~\ref{setup}.c), allowing for a just-in-time evaluation of physical processes, as students may directly observe and evaluate the process of heating and the formation of a stationary state with the help of these features. The smartglass setup using HoloLens at the same time ensures a hands-free scenario, in which students are able to interact with the physical setup of the experiment without losing track of the measurement data. Therefore, we expect this setting to intensify the theory-experiment interactions \cite{theoexpint}, since direct feedback in form of different representations to experimental actions of the users allows for immediate comparison to theoretical principles and predictions for further experimental reaction. Following the spatial contiguity principle of multimedia learning theory, the AR annotations thus are presented in direct proximity to the real objects to enhance learning effects \cite{Maye10} and facilitate memorization by the learners \cite{Fuji12}. 


\subsection{Theoretical background}
The paradigm experiment for heat conduction in metals can be realized with a metallic rod, which is heated on one end with a constant heating power, while being cooled on the other end \cite{Parr75}. If the rod is perfectly isolated, after some equilibration time the system will reach a steady state with a hot and a cold end and a constant spatial gradient along the rod axis, which allows to calculate the thermal conductivity constant $\lambda$ of the material. In general in one dimension---we choose $x$ along the axis of the rod---the temperature distribution $T(x,t)$ obeys the heat equation
\begin{equation}\label{heateq}
\frac{\partial}{\partial t}T(x,t)=\frac{\lambda}{\varrho c} \frac{\partial^2}{\partial x^2}T(x,t)
\end{equation}
with $\varrho$ being the density and $c$ the specific heat capacity of the material. According to Fourier's Law the heat flux $\dot{Q}$, i.e. the amount of energy that flows through an area $A$ per time, is given by 
\begin{equation}\label{fourier}
\dot{Q}=-\lambda\,A\,\frac{\partial}{\partial x}T(x,t).
\end{equation}
In the stationary state and in the case of perfect isolation, the constant heating power $\dot{Q}$ applied to the warm end of the rod will be removed on the cold end by cooling such that according to \eqref{fourier} a liner temperature gradient $(T_1-T_2)/L$, where $L$ is the lenght of the rod and $T_i$ are the two end-temperatures, is established. Thus, the thermal conductivity constant can be computed according to
\begin{equation}\label{lambd}
\lambda = \frac{L}{A\,(T_1-T_2)}\dot{Q},
\end{equation}
if the constant heating power $\dot{Q}$ and the dimensions, i.e. length $L$ and cross-section $A$, of the rod are known. This, however, only holds for a perfectly isolated rod. In the case of heat exchange with the environment, modeled as heat bath with constant temperature $T_0$, a Newtonian cooling term proportional to the temperature difference to the environment $T-T_0$ has to be added on the right-hand side of \eqref{heateq}, leading to an equation of the form\cite{Parr75} 
\begin{equation}\label{heateqh}
\frac{\partial}{\partial t}T(x,t)=\frac{\lambda}{\varrho c} \frac{\partial^2}{\partial x^2}T(x,t) -\frac{h}{cm}(T(x,t)-T_0),
\end{equation}
with $h$ being the heat transfer coefficient and $m$ the mass of the sample, which finally has the exponential solution \cite{Parr75}
\begin{equation}\label{expdec}
T(x,t) = T_0 + (T_1-T_0)\textrm{e}^{-\alpha x}
\end{equation}
for the temperature profile in this case. If the coefficient $\alpha$ is determined experimentally, the heat transfer coefficient can be calculated according to
\begin{equation}\label{heq}
h = \alpha^2 \lambda AL.
\end{equation}
Therefore, also the loss of heat to the environment can be determined experimentally.

\subsection{Experimental setup}\label{exp}

\begin{figure}[t]
 \begin{flushleft}
 \includegraphics[height=4.7cm]{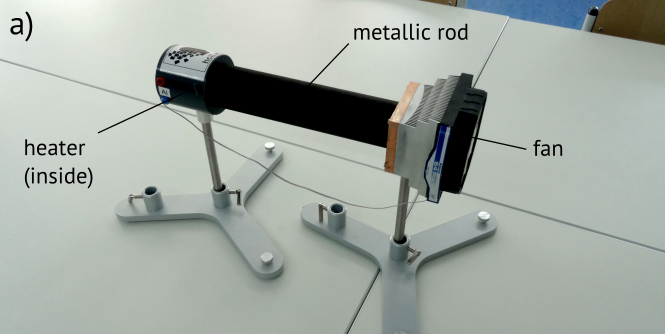}\,\includegraphics[height=4.7cm]{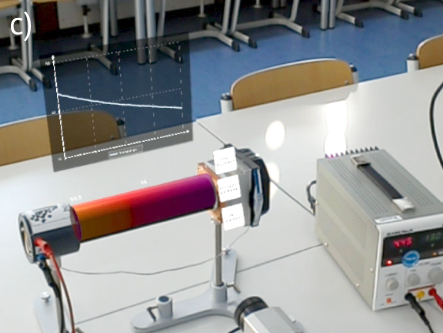}\\\vspace*{0.3mm}
 \includegraphics[height=4.7cm]{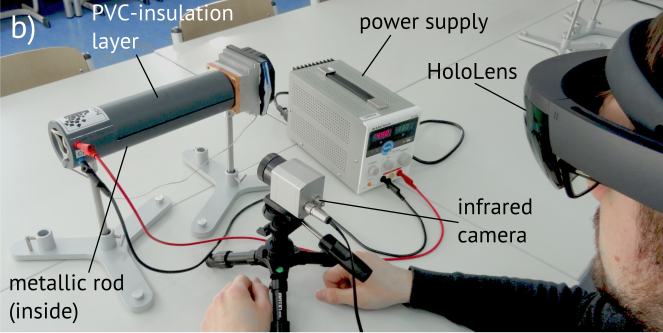}\,\includegraphics[height=4.7cm]{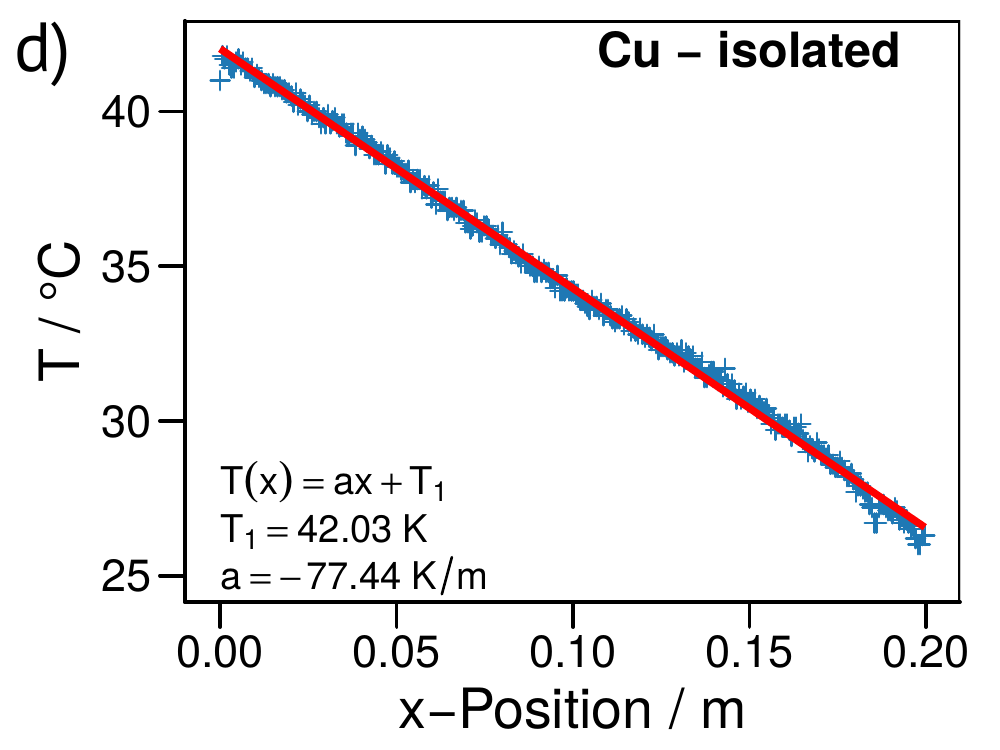}
 \end{flushleft}
 \caption{a) Experimental setup (non isolated rod); b) Experimental setup (rod with PVC-insulation) and user wearing a HoloLens; c) holo.lab setup (non isolated rod) with MR experience; augmented representations: false-color representation of temperature along the rod, numerical values at three points above the rod, temperature graph; d) Temperature graph after equilibration (blue crosses) and linear fit (red line) for an isolated Cu-rod.}
 \label{setup}
\end{figure}

The heat conduction experiment used in our laboratory courses consists of cylindrical metal sample rods with a length of $L=26$\,cm and a diameter of $d=5$\,cm, made of Aluminum and Copper, respectively (cf. Fig.~\ref{setup}.a). The sample is heated at one end with a cartridge heater and cooled at the other with a standard CPU fan. The isolated version moreover has a PVC insulation layer with a 3\,mm slit along the rod, to allow for thermal imaging of the rod inside (cf. Fig.~\ref{setup}.b). The temperature data finally is extracted from thermal images taken with an Optris PI 450 IR camera placed in front of the sample. The temperature values in axial direction are then passed to the HoloLens via WiFi, where the visualization is done. In the current state, the HoloLens-App developed by the authors provides three different representations of the data: false-color image of the temperature values, projected as ``hologram'' directly onto the sample cylinder, numerical values at three pre-defined points (the two ends and the middle of the rod) and a temperature graph as a function of the position along the rod, hovering above the setup, cf. Fig.~\ref{setup}.c. For this arrangement the spatial contiguity principle is not only obeyed for the false-color representation by default, but also for the numerical and the graph representation, as the $x$-position of the values exactly corresponds to the position in real space along the rod. Since the augmentations are moreover shown in direct neighbourhood to the experimental apparatus, both can be observed and compared at the same time. The user may switch the numerical and graph representation on and off at will; moreover, it is possible to export the data as csv-file at any time for later analysis. These functions can be executed with the help of virtual buttons---square holograms projected at the right end of the rod---which can be chosen with the so-called gaze point, a cursor, that can be moved with the users gaze. If the gaze point meets a hologram ready for selection, it will be highlighted. It may then be selected with the air-tap gesture, a hand gesture analogue of clicking on a mouse or a touch pad, simply by tapping with the forefinger \cite{gestures}. 

Since in many laboratory courses it is common for the students to work at the same experimental apparatus simultaneously in groups of two or even more, another important feature of the current design consists of the possibility to cooperate not only in the real, but also in the virtually augmented world. In the case of this setup this is ensured, as all students attending the experiment and wearing a HoloLens are able to see and to work with the same augmentations, enabling true cooperation also in MR. Therefore, they are not only able to discuss the experimental progress during the session, but also the virtual annotations and evaluations presented in their shared MR experience. 
This is possible since HoloLens itself guarantees a very high quality level of spatial registration of the virtual objects in real space. This means that a hologram, once positioned in space, stays there, even if the user moves around, leaves the room and comes back again. The initial positioning of the holograms in our setup is achieved with the help of a visual marker on the experimental apparatus: after successful marker recognition the AR content is simply presented in spatial relation to the position of the marker. If two or more students use a HoloLens, both devices use the marker recognition and present the AR content independent of each other.
The gesture recognition of HoloLens works in a limited region, the gesture frame, which is within easy reach of a person’s hands. Interference with a person performing the gestures in immediate neighbourhood is possible, however, in a normal working distance in the laboratory very unlikely and would only occur on purpose. That means, that actually all persons only control their own HoloLens.

Compared to projection-based AR techniques, which also are able to overcome the split attention effect, the holo.lab setting has the advantage of augmenting 3D objects without any occlusion issues that are always present otherwise. Moreover, additional augmentations---not fixed to surfaces of real objects---can be implemented here. One drawback of state-of-the-art smartglasses, however, is the limited field of view. While this is an issue especially for large-scale experiments, in table-top setups it represents only a minor restriction that will be overcome in the next generation of smartglasses.

The standard version of the experiment in the framework of the laboratory course, as used by the control group of our study (cf. Sec. \ref{study}), does not use the HoloLens and therefore cannot benefit from real-time augmentations and representations of the data. Here students instead have to use a hand-held FLIR i60 IR camera, a standard device in the laboratory, to take thermal images of the sample. These images may then be transferred to a provided PC, where temperature data can be extracted from them using the commercial software of the camera manufacturer. Only then the data can be exported to a csv-file for further processing and plotting. To identify the stationary state of the temperature distribution, this has to be done several times during the heating procedure. Therefore, the main differences of this setup compared to the HoloLens setup are that neither spatial nor temporal contiguity of the representations of the data are provided in this scenario: students do not get any kind of real-time feedback besides the false-color image. This also means that they do not get different representations of the experimental data at the same time, but first have to process the snapshots taken with the hand-held IR camera; finally displaying the data is also locally decoupled from the physical experimental setup.

\subsection{Experimental results}

In both setups the data produced during the laboratory course has afterwards to be processed and analyzed by the students. Extracting the slope of a linear fit to the temperature distribution of the stationary state of the isolated rods (cf. Fig.~\ref{setup}.d) is sufficient to calculate the thermal conductivity $\lambda$ of the materials in use via \eqref{lambd}. Such an analysis  of the data is also part of the laboratory course and has to be carried out by all of the students after the experimental phase to analyze and discuss the results and difficulties of the experiment. Regarding the very simple realization of the experimental setup---especially the non perfect PVC insulation of the rods and the cooling against room temperature with the fan---at a constant heating power of $\dot{Q} = 50\,{\rm W}$ the setup still yields reasonable results for the thermal conductivity constant $\lambda_{\rm Al}=(128\pm 11)\,{\rm Wm}^{-1}{\rm K}^{-1}$ and $\lambda_{\rm Cu}=(329\pm 27)\,{\rm Wm}^{-1}{\rm K}^{-1}$ of Aluminum and Copper, respectively, as already has been reported in \cite{Strz17}. These values underestimate the reference values found in literature ($\lambda_{\rm Al, lit}= 235\,{\rm W/(m\,K)}$ and $\lambda_{\rm Cu, lit}= 401\,{\rm W/(m\,K)}$), since thermal losses to the environment effectively reduce the thermal flux through the rods. The topic of these losses is then again picked up by the analysis of the experiment using the non isolated rods. Extracting the decline factor $\alpha$ from an exponential fit to the data  of the form \eqref{expdec} allows to calculate the heat transfer coefficient $h$ according to \eqref{heq}, yielding $h_{\rm Al} = (0.72 \pm 0.01)\,{\rm WK}^{-1}$ and $h_{\rm Cu} = (2.5 \pm 0.01)\,{\rm WK}^{-1}$, respectively. Expectedly these values are again underestimating the reference values $h_{\rm Al,lit} = (1.32 \pm 0.01)\,{\rm WK}^{-1}$ and $h_{\rm Cu,lit} = (3.05 \pm 0.01)\,{\rm WK}^{-1}$,  where the literature values for $\lambda$ were used. Nevertheless they are reasonable regarding the very simple insulation of the setup. For the analysis of the data performed by the students, a thorough discussion of the shortcomings of the setup and the data collection are required. In this context the issue of heat losses to the environment also alludes to technical aspects as well as the principle of energy conservation and different types of heat transport, such as radiation and convection and therefore is an important part of the subjects of this experimental course.


\section{Empirical study on conceptual understanding}\label{study}

\subsection{Theoretical background}

The benefit of the holo.lab setup presented in this paper is the possibility of keeping track of the real physical devices and representations of the data simultaneously and in real-time. The false-color representation allows to experience an otherwise invisible quantity, like in this case temperature, with human senses, thus extending perception to new regimes \cite{Voll01,Voll13,Hagl16a,Palm16,Strz17}. Based on multimedia learning theory \cite{Maye10}, it can be assumed that temporal and spatial contiguity of the holographic projection directly onto the real object effectively supports the learning process by avoiding the split attention effect that would occur according to CLT \cite{Swel10}, if other display types, like tablets or even computer screens, would be used for the virtual augmentation. In fact, basic principles of multimedia learning theory can simply be extended to the case of AR scenarios if real objects resemble pictures, while virtual augmentations substitute written text \cite{Sant14}. Therefore, real world annotations ensure spatial contiguity and reduce the cognitive load of the students \cite{Sant14} such that a larger fraction of the short term memory may be used during the cognitive process according to CLT \cite{Swel10}. This is also supported by user studies targeting the ability to memorize annotated information in AR settings \cite{Fuji12}; also here spatial contiguity, i.e.~presenting the information directly on top of an object, turned out to significantly improve the results. Additionally the MR setup automatically implements a contextual visualization \cite{Craw01} of the data by design \cite{Sant13,Sant14}. In combination with the real-time presentation of the data, we thus expect a more effective interrelation of theory and experiment, as, e.g., the equilibration process can directly be followed watching the temperature graph or the false-color representation and be related to theoretical predictions, which otherwise---in classical laboratory class setups---is only possible in hindsight while analyzing experimental data outside the lab. Therefore, in this design a direct feedback is implemented, ensuring that students get an immediate impression of the effects of the experimental parameters. Besides this direct feedback, competent handling of representations is a key to successful physics learning \cite{Cock12,Ains99,Heuv91,Niem12,Melt05,Kohl08,Klei17} and so the integrative use of multiple representations (such as diagrams, measuring values and true- or false-color images) should foster conceptual understanding. Moreover, in this setting students are also able to observe intermediate non-equilibrium states and their evolution to the final equilibrium state. 

\subsection{Material and methods}

In order to analyze the effects of the MR enhancements of the experiment on the conceptual understanding of the students concerning heat and temperature, we thus conducted an empirical study in the summer term 2017 at University of Kaiserslautern paralleling the introductory laboratory courses in physics. The overall number of $N = 59$ participants (average age of 20.7 years; almost all of them were in their second term of study) can be divided into $N_{\rm C} = 43$ students in the control group (CG) and $N_{\rm T} = 16$ students in the treatment group (TG). The CG performed the experiment in the framework of the laboratory course working with the traditional, well-established experimental setup equipped with a hand-held IR camera and a PC---without a just-in-time analysis of the data collected---while the participants of the TG used the holo.lab setup with the HoloLens in parallel to the standard laboratory course as described in Sec.~\ref{exp}. Our evaluation in a 2$\times$2-design was integrated into the standard processes of the laboratory course:
 After arrival at the lab, students first have to pass a short introductory test, consisting of few oral questions by their advisor, granting that they have the background knowledge required to use the experimental apparatus. After that, the TG additionally was instructed to use the HoloLens including a short calibration of the device. The concept test was conducted in a pre- and post-experiment design. Before all of the students (TG and CG) started to experiment, they first had to complete the pre-test; having finished their experimental work they completed the post-test. The time slot for the laboratory course is set to three hours, such that after all more than two hours of pure experimenting time are available for the participants.  

The items of our conceptual evaluation were picked from the Heat and Temperature Conceptual Evaluation (HTCE) \cite{HTCE} which is a multiple choice survey consisting of 28 items dealing with basic concepts of thermodynamics and statistics at an introductory college level, which had to be translated into German. In a literature review concerning concept tests for thermodynamics the HTCE turned out to cover the topics, which are relevant in our case, best. Its items can be clustered into eight groups covering different conceptions \cite{Tana06}. From these we have chosen 16 items, covering the scope of the experiment, i.e.~the clusters Heat and Temperature (Items 1-4), Rate of Cooling (Items 5-7), Rate of Heat Transfer (Items 10, 11), Specific heat capacity (Items 16-19) and Thermal conductivity (Items 26-28). The survey was presented to the students on tablet computers using an online assessment tool.

\subsection{Results}

\begin{figure}[t!]
 \begin{center}
\includegraphics[scale=.75]{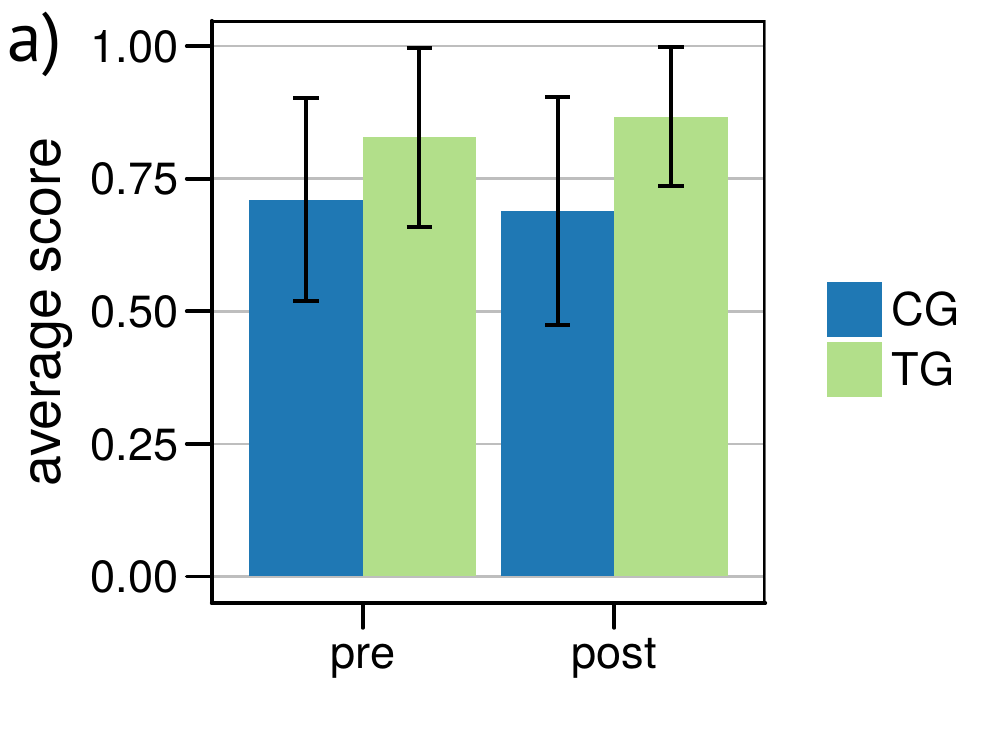}\includegraphics[scale=.75]{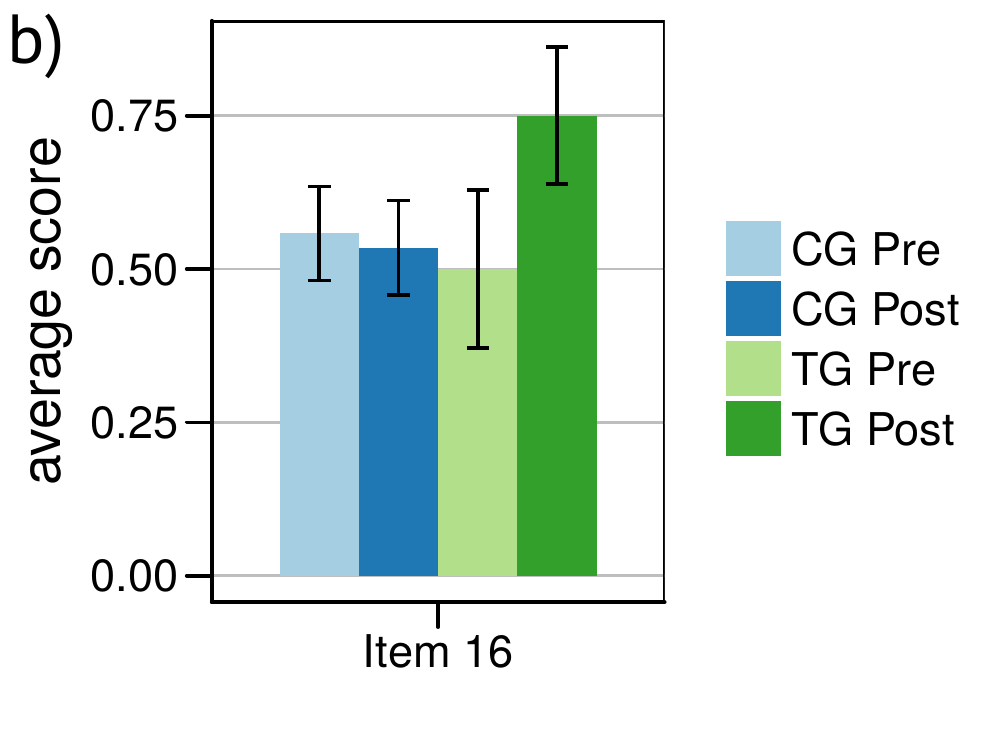}
 \end{center}
 \caption{a) Average pre- and post-scores of the concept test (HTCE) of CG and the TG, respectively. The errorbars indicate the standard deviaion; b) Average pre- and post-scores of Item 16 of CG and the TG, respectively. The errorbars indicate the standard error.}
 \label{scores}
\end{figure}

First we checked the reliability of the test, achieving an acceptable value of Cronbach's $\alpha = 0.76$. A Kolmogorov-Smirnov test of the scores of the participants  showed agreement with a normal distribution. However, we found differences in the pre-knowledge between the two groups, the TG on average reaching higher scores compared to the CG ($t$-test statistics: $t = 2.29$, $df = 30$, $p = 0.029$, Cohen's $d = 0.63$), cf.~Fig.~\ref{scores}.a. 

To analyze the data for statistically significant gains in the conceptual understanding of the two groups, we performed $t$-tests and calculated the corresponding effect size. The CG did not show a significant effect ($t$-test statistics: $t = 1.19$, $df = 42$, $p = 0.242$), whereas participants of the TG on average could improve their score ($t$-test statistics: $t = 2.30$, $df = 15$, $p = 0.036$, Cohen's $d = 0.26$), cf.~Fig.~\ref{scores}.a.

We compared the post-test scores of both groups using an analysis of covariance (ANCOVA) with the pretest-scores as covariates to take the different pre-knowledge into account. This yields a statistically significant ($F = 5.61, p = 0.02$) corresponding small effect size with a tendency towards the medium range (adjusted Cohen's $d = 0.43$). We therefore could conclude, that over all the MR-setting of the TG improves students' gain in conceptual understanding in the context of heat transfer and thermal conduction. 

A more detailed analysis based on single items
shows that the score gain for Item 16 (cf. Appendix) is especially high compared to all of the other items. Here the TG is able to increase the score by 0.25 from 0.5 to 0.75, corresponding to a Hake's normalized gain of $\langle g_{\rm H}\rangle = 0.5$, cf.~Fig.~\ref{scores}.b. Taking into account that this item also belongs to the group of items with the lowest pre-scores this is a remarkable result. In Item 16 students have to qualitatively identify a connection between a constant heat flow into an otherwise perfectly isolated system and the graph representation of the temperature of the system. Exactly this kind of diagram (temperature graph over time) is not part of the augmented representations the TG can benefit from, however, the real-time feedback allows to easily follow the development of the temperature during the heating process, which is not possible for the CG where only pictures with the IR-camera for later analysis are taken. Therefore, this result might be a first hint that augmentations with real-time feedback could enhance the interplay between experiments and theoretical representations.

\section{Conclusion}

In this paper we presented a first holo.lab setting for augmenting an existing experimental setup for the analysis of thermal conductivity of metals, well known in physics laboratory courses. Making use of the capabilities of state-of-the-art smartglasses opens the possibility of visualizing invisible physical quantities. In the scenario shown here, we extend human perception to new regimes---temperature and heat---by digitally augmenting physical objects with their own temperature distribution in false-color representation in real-time. Based on several theories, we could expect such a holo.lab setting to support the learning process and to strengthen the connection between theory and experiment. The fixation on the real object provides an intrinsic contextuality \cite{Craw01}, while at the same time students have their hands free to use the just-in-time evaluation of the data to directly examine the physical processes and parameters involved, and immediately compare the outcome to theoretical predictions, which we believe to enhance the links between theory and experiment. Moreover, spatial and temporal contiguity principles of CATML are obeyed, simultaneously avoiding a split-attention effect of aCLT, which is supposed to support the learning process of the students \cite{Maye10}. 


First positive results of the evaluation of the conceptual understanding of students support our belief in the benefits of AR and encourage us to continue the development of the holo.lab with smartglasses for the general use in laboratory courses in physics. Even though the added value of AR was very limited in this example (in order to establish a fair comparison between groups), we found positive results. Hence, we can assume that more complex experiments will benefit even more from augmentation. This will be tested in upcoming scenarios. 
Moreover, further evaluations have certainly to be carried out capturing and analyzing affective and cognitive variables of the participants, especially cognitive load, to validate our assumptions. The real-time interaction with different representations as well as the corresponding impact on the conceptual and representational understanding of the students will also need further examination. Moreover, also the aspect of cooperation of the students in the framework of holo.lab experiments and the exact interactions of theory and experiment will be analyzed in future studies.

The basic idea of a holo.lab scenario is to augment existing standard experiments widely used in laboratory classes in Physics. This has the advantage of being easily transferable also to other laboratories at different universities. However, the programming effort for the integration of new sensors, e.g., different IR cameras, into the software framework, as well as the relatively high price for the Development Edition of HoloLens \cite{holoprice} are limiting factors for a rapid implementation of holo.lab settings in different laboratories at the moment.

\subsection*{Acknowledgement:}
Support from the German Federal Ministry of Education and Research (BMBF) via the projects \textit{Be-greifen} \cite{begreifen} and gLabAssist is gratefully acknowledged.
Furthermore we thank our colleagues Prof.~Dr.~Artur Widera, Dr.~Britta Leven and Bernd Stabel for intensive discussion in this context.

\pagebreak

\section{Appendix}

\subsection*{HTCE Item 16:}

This question refers to a cup which contains water at room temperature. The cup is perfectly insulated so that no heat can transfer into or out of the cup. A small coffee cup heater inside the cup is used to transfer heat to the water. The water does not boil. 

If heat is transferred to the cup at a steady rate, which of the graphs below best represents the shape of the graph of the temperature of the water over time as the heat is transferred? Answer G if you think that none is correct.\\

\begin{figure}[h!]

A)\includegraphics[width=0.3\linewidth]{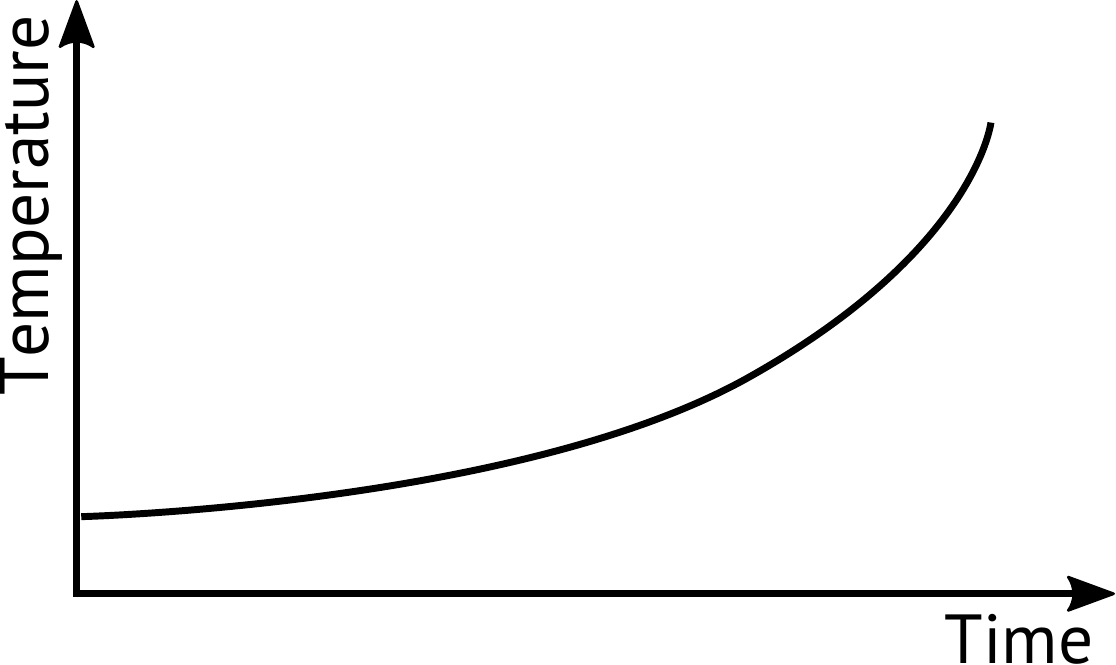}
B)\includegraphics[width=0.3\linewidth]{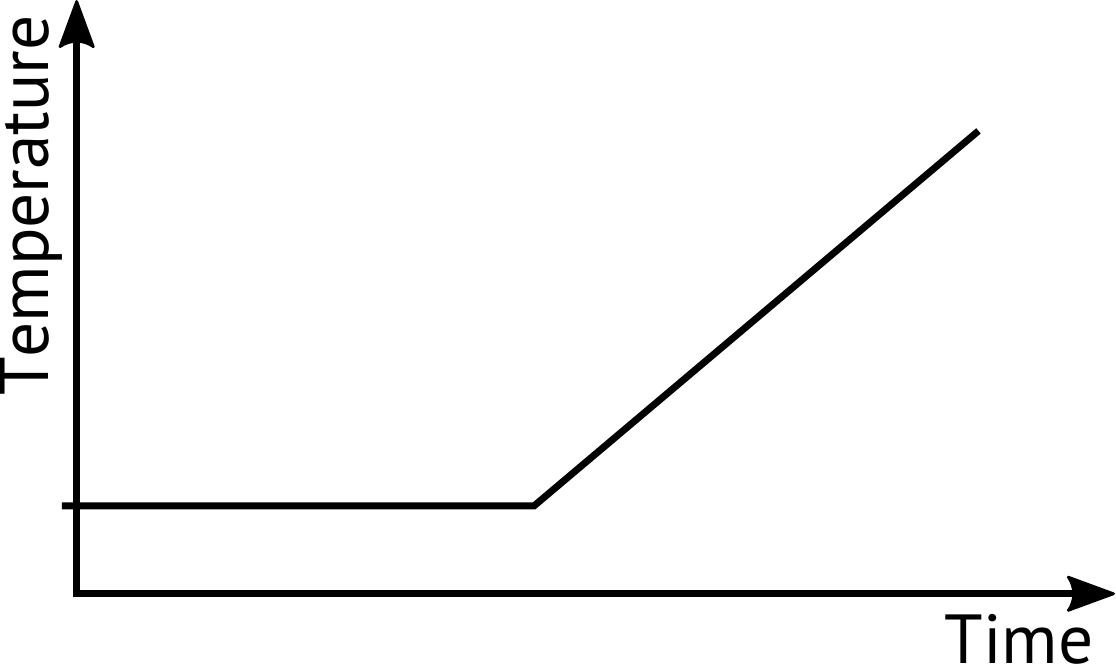}
C)\includegraphics[width=0.3\linewidth]{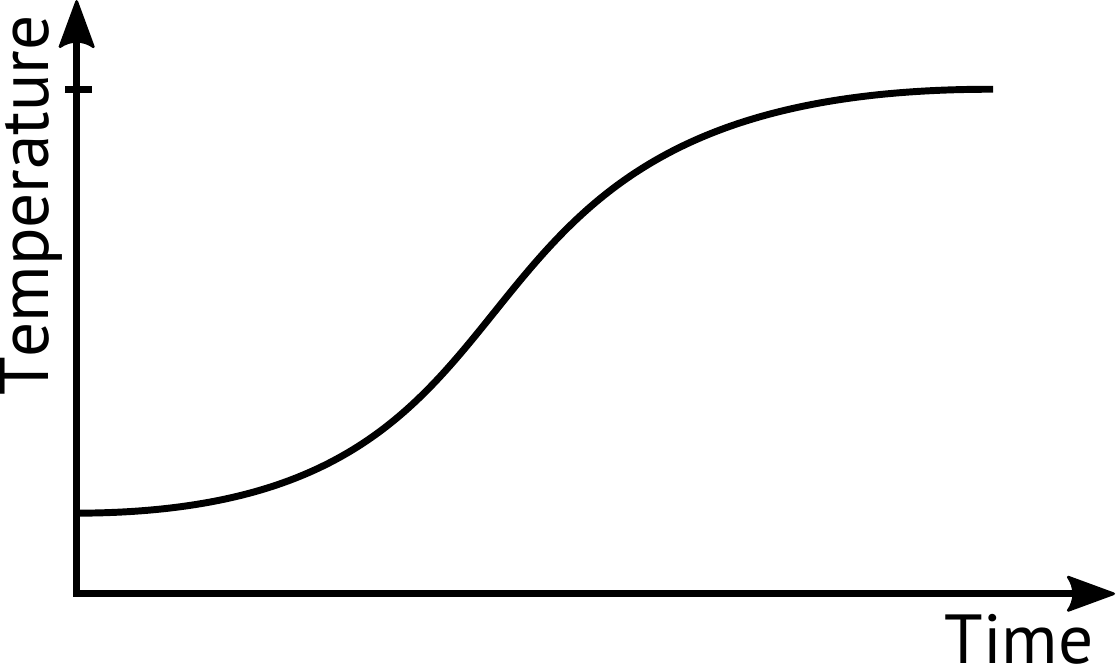}\\

D)\includegraphics[width=0.3\linewidth]{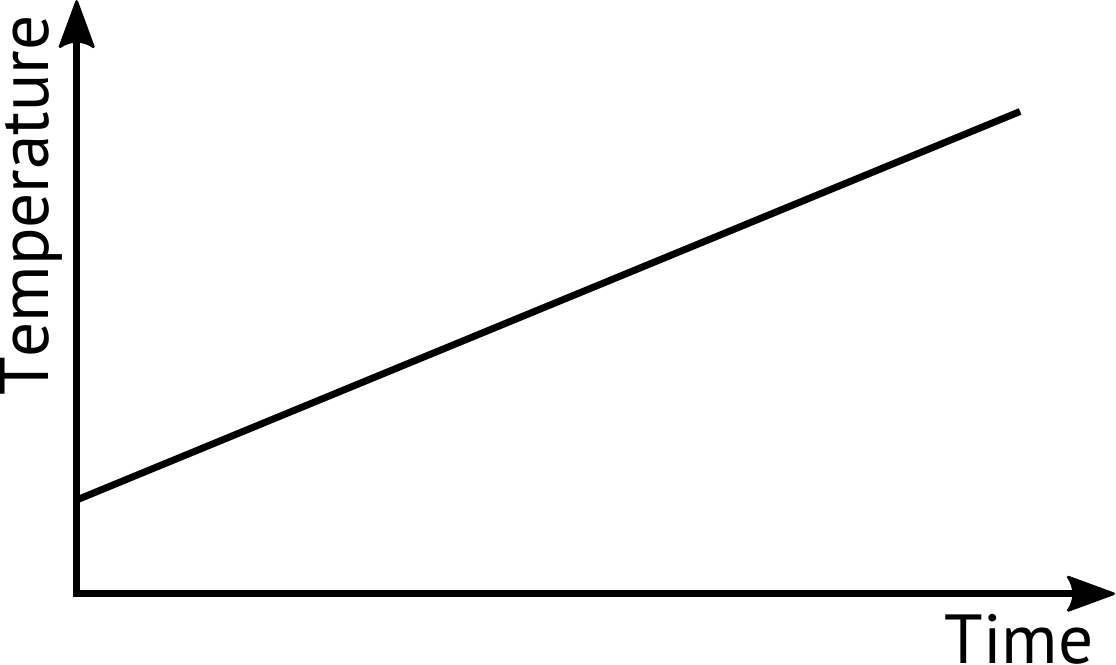}
E)\includegraphics[width=0.3\linewidth]{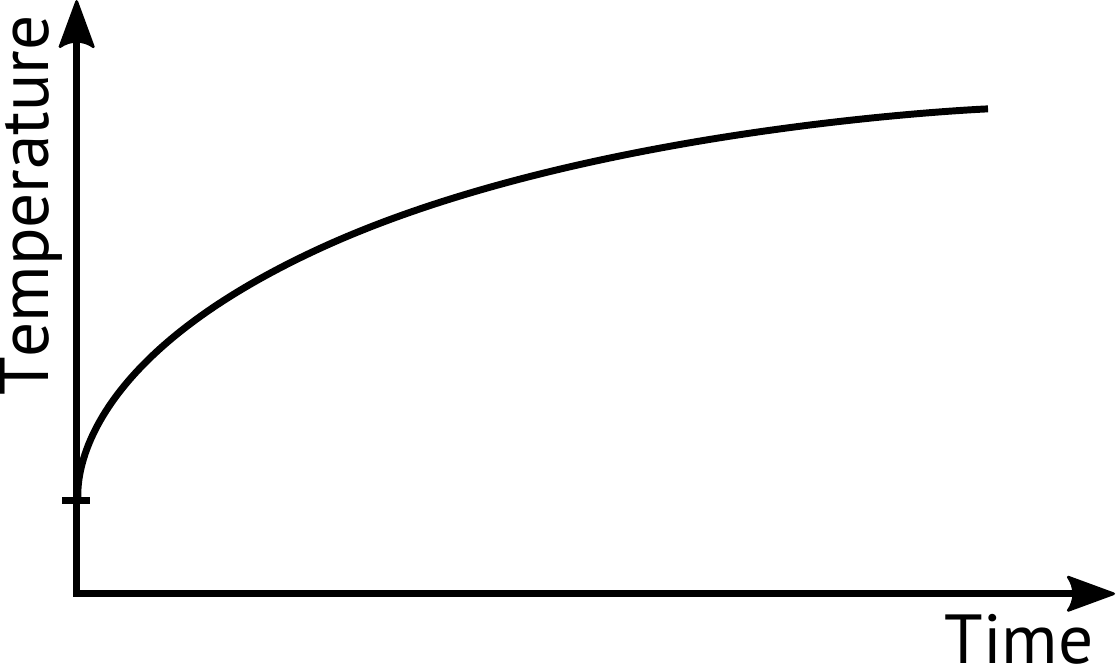}
F)\includegraphics[width=0.3\linewidth]{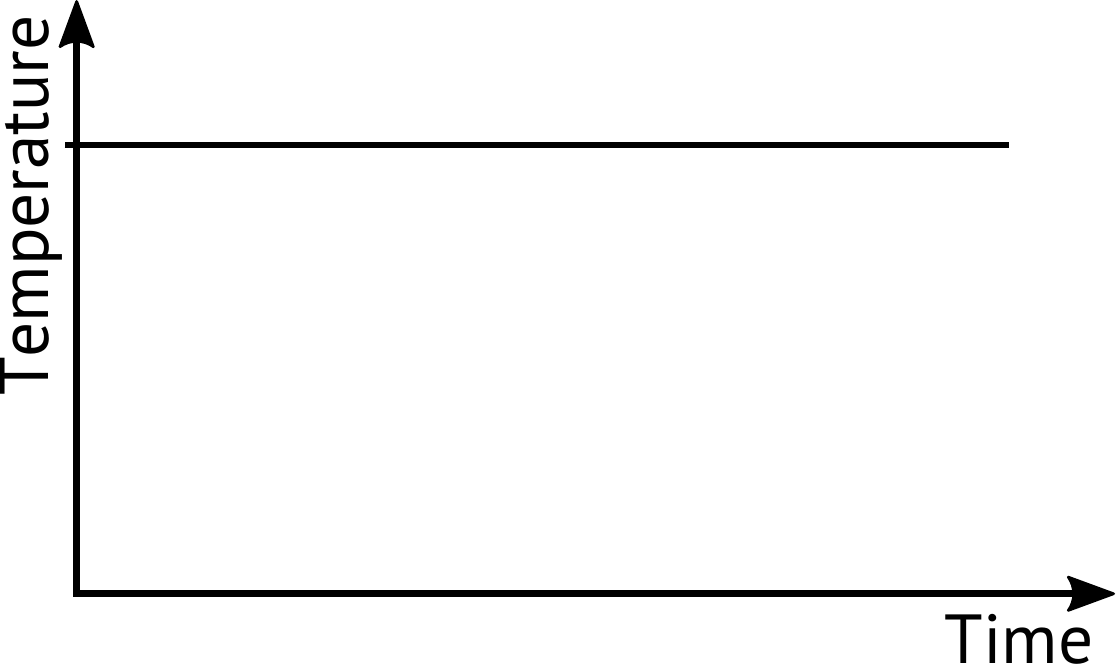}
\end{figure}

\section*{References}

\bibliographystyle{unsrt}


\begin{thebibliography}{10}

\bibitem{Klei15}
P.~Klein, J.~Kuhn, A.~Müller, and S.~Gröber.
\newblock Video analysis exercises in regular introductory mechanics physics
  courses: Effects of conventional methods and possibilities of mobile devices.
\newblock In W.~Schnotz, A.~Kauertz, A~H.~Ludwig, Müller, and J.~Pretsch,
  editors, {\em Multidisciplinary Research on Teaching and Learning}, pages
  270--288. Palgrave Macmillan, 2015.

\bibitem{Kuhn14}
J.~Kuhn, A.~Molz, S.~Gröber, and J.~Frübis.
\newblock iRadioactivity - Possibilities and Limitations for Using Smartphones
  and Tablet PCs as Radioactive Counters.
\newblock {\em Phys. Teach.}, 52:351--356, 2014.

\bibitem{Schw13}
O.~Schwarz, P.~Vogt, and J.~Kuhn.
\newblock Acoustic measurements of bouncing balls and the determination of
  gravitational acceleration.
\newblock {\em Phys. Teach.}, 51:312--313, 2013.

\bibitem{Vogt11}
P.~Vogt, J.~Kuhn, and S.~Müller.
\newblock Experiments using cell phones in physics classroom education: The
  computer aided g-determination.
\newblock {\em Phys. Teach.}, 49:383--384, 2011.

\bibitem{Klei14}
P.~Klein, M.~Hirth, S.~Gröber, J.~Kuhn, and A.~Müller.
\newblock Classical experiments revisited: Smartphone and Tablet PC as
  experimental Tools in Acoustics and Optics.
\newblock {\em Phys. Educ.}, 49:412--418, 2014.

\bibitem{Kuhn14a}
J.~Kuhn.
\newblock Relevant information about using a mobile phone acceleration sensor
  in physics experiments.
\newblock {\em Am. J. Phys.}, 82:94, 2014.

\bibitem{Maye10}
R.~E. Mayer, editor.
\newblock {\em The Cambridge Handbook of Multimedia Learning}.
\newblock Cambridge University Press, 2010.

\bibitem{Swel10}
J.~Sweller.
\newblock Implications of cognitive load theory for multimedia learning.
\newblock In {\em The Cambridge Handbook of Multimedia Learning}. Cambridge
  University Press, 2010.

\bibitem{Huk09}
T.~Huk and S.~Ludwigs.
\newblock Combining cognitive and affective support in order to promote
  learning.
\newblock {\em Learning and Instruction}, 19(6):495 -- 505, 2009.

\bibitem{More07}
R.~Moreno and R.~Mayer.
\newblock Interactive multimodal learning environments.
\newblock {\em Educ Psychol Rev}, 19:309--326, 2007.

\bibitem{Schm16}
D.~Schmalstieg and T.~Höllerer.
\newblock {\em Augmented Reality: Principles and Practice}.
\newblock Addison-Wesley Professional, 2016.

\bibitem{Sand15}
C.~Sandor, M.~Fuchs, {\'{A}}.~Cassinelli, H.~Li, R.~A. Newcombe, G.~Yamamoto,
  and S.~K. Feiner.
\newblock Breaking the barriers to true augmented reality.
\newblock {\em CoRR}, abs/1512.05471, 2015.

\bibitem{Hock16}
P.~Hockett and T.~Ingleby.
\newblock Augmented reality with hololens: Experiential architectures embedded
  in the real world.
\newblock {\em CoRR}, abs/1610.04281, 2016.

\bibitem{Sant14}
M.~E.~C. Santos, A.~Chen, T.~Taketomi, G.~Yamamoto, J.~Miyazaki, and H.~Kato.
\newblock Augmented reality learning experiences: Survey of prototype design
  and evaluation.
\newblock {\em IEEE Transactions on Learning Technologies}, 7(1):38--56, Jan
  2014.

\bibitem{Kuhn16}
J.~Kuhn, P.~Lukowicz, M.~Hirth, A.~Poxrucker, J.~Weppner, and J.~Younas.
\newblock gPhysics -- Using Smart Glasses for head-centered, context-aware
  learning in Physics Experiments.
\newblock {\em IEEE Transactions on Learning Technologies}, 9(4):304--317, Oct
  2016.

\bibitem{Milg94}
P.~Milgram and F.~Kishino.
\newblock A taxonomy of mixed reality visual displays.
\newblock {\em IEICE Trans. Information and Systems}, 77(12):1321--1329, 1994.

\bibitem{Voll01}
M.~Vollmer, K.-P. Möllmann, F.~Pinno, and D.~Karstädt.
\newblock There is more to see than eyes can detect.
\newblock {\em The Physics Teacher}, 39(6):371--376, 2001.

\bibitem{Moel07}
K.-P. Möllmann and M.~Vollmer.
\newblock Infrared thermal imaging as a tool in university physics education.
\newblock {\em European Journal of Physics}, 28(3):S37--S50, 2007.

\bibitem{Voll13}
M.~Vollmer and K.-P. Möllmann.
\newblock {\em Infrared Thermal Imaging}.
\newblock Wiley, Weinheim, 2013.

\bibitem{Hagl16}
J.~Haglund, F.~Jeppsson, and K.~J. Schönborn.
\newblock Taking on the heat---a narrative account of how infrared cameras
  invite instant inquiry.
\newblock {\em Research in Science Education}, 46(5):685--713, Oct 2016.

\bibitem{Hagl16a}
J.~Haglund, F.~Jeppsson, E.~Melander, A.-M. Pendrill, C.~Xie, and K.~J.
  Schönborn.
\newblock Infrared cameras in science education.
\newblock {\em Infrared Physics \& Technology}, 75:150 -- 152, 2016.

\bibitem{Nord16}
J.~Nordine and S.~Weßnigk.
\newblock Exposing hidden energy transfers with inexpensive thermal imaging
  cameras.
\newblock {\em Science Scope}, 39(7):25--31, 2016.

\bibitem{Palm16}
K.~L. Palmerius and K.~Schönborn.
\newblock {\em Visualization of Heat Transfer Using Projector-Based Spatial
  Augmented Reality}, pages 407--417.
\newblock Springer International Publishing, Cham, 2016.

\bibitem{theoexpint}
The theoretical knowledge of the students enables them to formulate predictions
  about the outcome of the experiments. While performing experiments, these
  predictions can be verified or falsified by the real measurement data, which
  might either lead to a confirmation of the theoretical model the students
  have in mind, or show inconsistencies with it. Contradictions to the
  theoretical model may in turn trigger a critical re-examination of the
  theoretical principles and the expected predictions resulting from them. This
  process usually is called interactions between theory and experiment.

\bibitem{Fuji12}
Y.~Fujimoto, G.~Yamamoto, H.~Kato, and J.~Miyazaki.
\newblock Relation between location of information displayed by augmented
  reality and user's memorization.
\newblock In {\em Proceedings of the 3rd Augmented Human International
  Conference}, AH '12, pages 7:1--7:8, New York, NY, USA, 2012.

\bibitem{Parr75}
J.~E. Parrot and A.~D. Stuckes.
\newblock {\em Thermal Conductivity of Solids}.
\newblock Pion Limited, London, 1975.

\bibitem{gestures}
More information about the use of gestures with HoloLens can be found in the HoloLens-support:
\newblock Use gestures,
\newblock https://support.microsoft.com/en-us/help/12644/hololens-use-gestures.
Alternatively also a clicker can be used, cf. 
\newblock Use the HoloLens clicker,
\newblock https://support.microsoft.com/en-us/help/12646/hololens-use-the-hololens-clicker.
Besides gestures and clicker, also speech recognition, which is not used in the setup presented in this paper, can be used to interact with HoloLens, cf. 
\newblock Use your voice with HoloLens, 
\newblock https://support.microsoft.com/en-us/help/12647/hololens-use-your-voice-with-hololens

\bibitem{Strz17}
M.~P. Strzys, S.~Kapp, M.~Thees, J.~Kuhn, P.~Lukowicz, P.~Knierim, and
  A.~Schmidt.
\newblock Augmenting the thermal flux experiment: A mixed reality approach with
  the hololens.
\newblock {\em The Physics Teacher}, 55(6):376, 2017.

\bibitem{Craw01}
M.L. Crawford.
\newblock Teaching contextually: Research, rationale, and techniques for
  improving student motivation and achievement in mathematics and science.
\newblock {\em CORD Comm}, 2001.

\bibitem{Sant13}
M.~E.~C. Santos, A.~Chen, M.~Terawaki, G.~Yamamoto, T.~Taketomi, J.~Miyazaki,
  and H.~Kato.
\newblock Augmented reality x-ray interaction in k-12 education: Theory,
  student perception and teacher evaluation.
\newblock In {\em 2013 IEEE 13th International Conference on Advanced Learning
  Technologies}, pages 141--145, July 2013.

\bibitem{Cock12}
M.~De~Cock.
\newblock Representation use and strategy choice in physics problem solving.
\newblock {\em Phys. Rev. ST Phys. Educ. Res.}, 8:020117, Nov 2012.

\bibitem{Ains99}
S.~Ainsworth.
\newblock The functions of multiple representations.
\newblock {\em Computers \& Education}, 33(2):131 -- 152, 1999.

\bibitem{Heuv91}
A.~Van Heuvelen.
\newblock Learning to think like a physicist: A review of research-based
  instructional strategies.
\newblock {\em American Journal of Physics}, 59(10):891--897, 1991.

\bibitem{Niem12}
P.~Nieminen, A.~Savinainen, and J.~Viiri.
\newblock Relations between representational consistency, conceptual
  understanding of the force concept, and scientific reasoning.
\newblock {\em Phys. Rev. ST Phys. Educ. Res.}, 8:010123, May 2012.

\bibitem{Melt05}
D.~E. Meltzer.
\newblock Relation between students’ problem-solving performance and
  representational format.
\newblock {\em American Journal of Physics}, 73(5):463--478, 2005.

\bibitem{Kohl08}
P.~B. Kohl and N.~D. Finkelstein.
\newblock Patterns of multiple representation use by experts and novices during
  physics problem solving.
\newblock {\em Phys. Rev. ST Phys. Educ. Res.}, 4:010111, Jun 2008.

\bibitem{Klei17}
P.~Klein, A.~M\"uller, and J.~Kuhn.
\newblock Assessment of representational competence in kinematics.
\newblock {\em Phys. Rev. Phys. Educ. Res.}, 13:010132, Jun 2017.

\bibitem{HTCE}
R.~K Thronton and D.~R. Sokoloff.
\newblock Heat and temperature conceptual evaluation.
\newblock https://www.physport.org/assessments/assessment.cfm?I=16\&A=HTCE,
  2001.

\bibitem{Tana06}
C.~Tanahoung, R.~Chitaree, C.~Soankwan, Sharma M., and I~Johnston.
\newblock Surveying thai and sydney introductory physics students’
  understandings of heat and temperature.
\newblock In {\em Proceedings of the Assessment in Science Teaching and
  Learning Symposium}, pages 190--194, 2006.

\bibitem{holoprice}
The current price of the Development Edition of HoloLens is 3000 US dollars.

\bibitem{begreifen}
https://www.physik.uni-kl.de/en/kuhn/forschungsprojekte/aktuelle-projekte/be-greifen/

\end{thebibliography}

\end{document}